\documentclass[12pt,preprint]{aastex}
\usepackage{graphicx}
\usepackage{epsfig}
\usepackage{rotating,colordvi}
\newcommand \ltw{\>\hbox{\lower.25em\hbox{$\buildrel <\over\sim$}}\>}
\shorttitle{Radio Emission Signatures in the Crab Pulsar}
\shortauthors{Hankins, Eilek}

\begin{document}
\title{RADIO EMISSION SIGNATURES IN THE CRAB PULSAR}
\author{T. H. Hankins\altaffilmark{1}}
\author{J. A. Eilek\altaffilmark{1}}
\altaffiltext{1}{Physics Department, New Mexico Tech, Socorro, NM 87801}
\email{thankins@aoc.nrao.edu}
\email{jeilek@aoc.nrao.edu}

\begin{abstract}

Our high time resolution observations of individual  pulses from
the Crab pulsar show that both the time and frequency signatures of
the interpulse are distinctly different from those of the main pulse.
Main pulses can occasionally be resolved into short-lived,
relatively narrow-band nanoshots.  We believe these nanoshots are
produced by soliton collapse in strong plasma turbulence.  Interpulses
at centimeter wavelengths are very different.   Their
dynamic spectrum contains regular, microsecond-long emission bands. We
have detected these bands, proportionately spaced in frequency, from
4.5 to 10.5 GHz. 
The bands cannot easily be explained by any current theory of pulsar
radio emission; we speculate on possible new models. 
\end{abstract}

\keywords{pulsars: individual (Crab Nebula pulsar) --- radiation
  mechanisms: non-thermal} 

\section{INTRODUCTION}

\label{section_Introduction}

What is the pulsar radio emission mechanism? Does the same mechanism
always operate, in all stars or throughout the magnetosphere of  one
star? What are the physical conditions in the magnetosphere 
that allow the emission to happen?   Despite forty years
of  effort, these questions still have not been answered
conclusively. 

Most models of pulsar radio emission fall into three groups. These are 
(1) antenna-type emission from coherent charge bunches;   (2) strong plasma
turbulence (SPT), in which nonlinear effects convert plasma waves to
electromagnetic waves which can escape the plasma;  and (3) several
variants of maser emission.  In Hankins \emph{et al.}\ (2003;  ``HKWE'') we
suggested these different emission mechanisms can be differentiated by
their time signatures, because the characteristic variability
timescales of each model differ.  To test this idea,  we designed and
developed data acquisition systems to probe the radio emission
signatures at the highest possible time resolutions. At low radio
frequencies scattering by electron density inhomogeneities in the
Crab Nebula and the interstellar medium mask the highest time and frequency 
resolution structure of the pulsar emission. The observations we
describe here were made at high enough frequencies to avoid the
pulse broadening, due to  multipath propagation through the
interstellar medium, which occurs at lower frequencies. 

\subsection{The Crab Pulsar}

\begin{figure}[htb]
\begin{center}
% Original bounding box: 72 72 540 720
%\includegraphics[width=1.20\columnwidth,viewport=54 0 540
%  640,clip]{f1}
%\includegraphics[width=0.8\textwidth]{f1.eps}
%\figurenum{1}
\epsscale{0.74}
\plotone{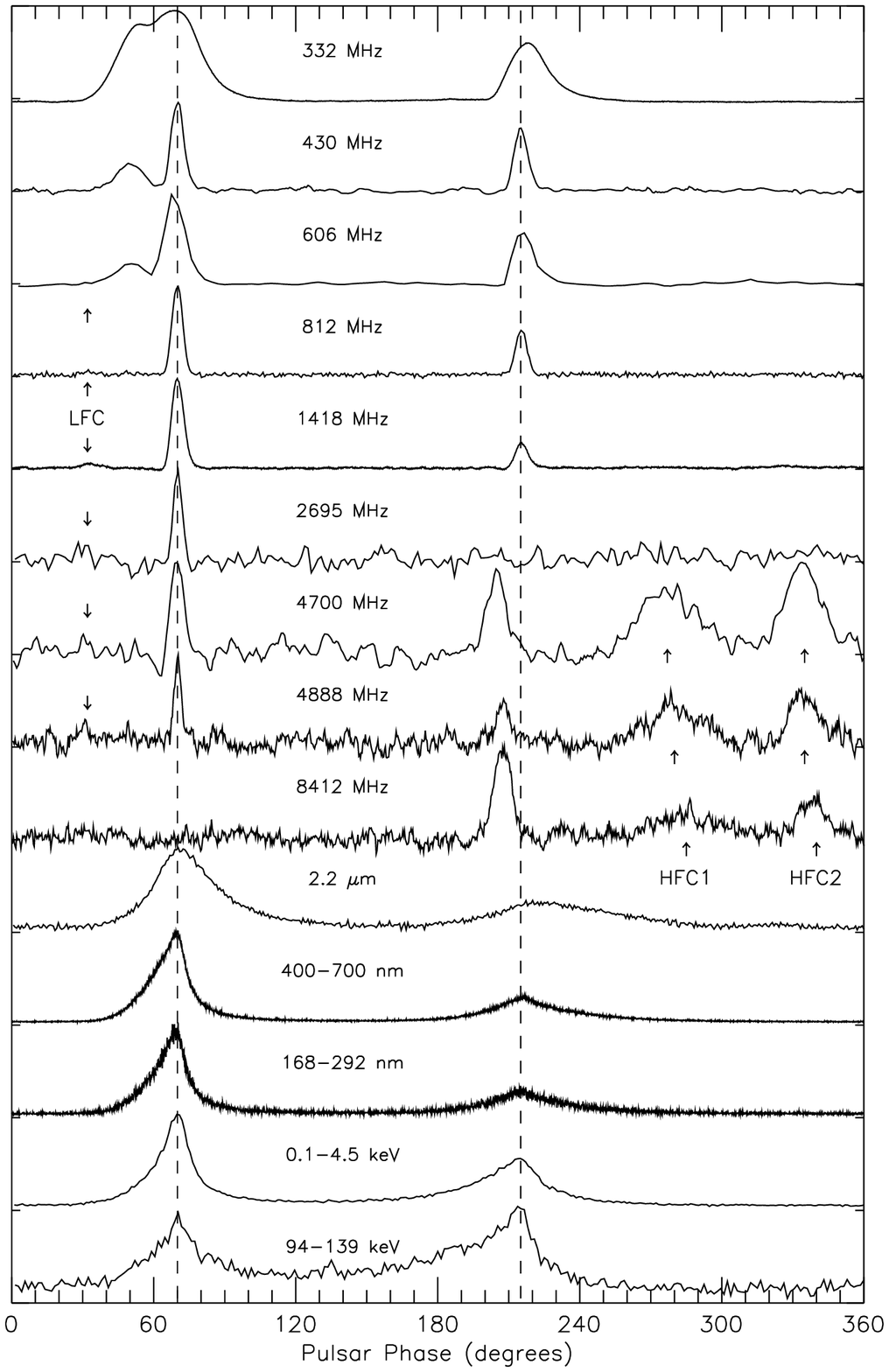}
\caption{The mean profile of the Crab pulsar, over a wide range of
frequencies (from Moffett \& Hankins 1996).   The main pulse and 
interpulse, shown by dashed lines at pulse phases $70^{\circ}$ and
$215^{\circ}$, persist from radio to hard X-ray bands.  However,
between 4.7 and 8.4 GHz the interpulse is offset from the interpulse
at lower and higher frequencies, and new components appear (labeled HFC1 and
HFC2). These intermediate-frequency components may have a different
origin from the lower and higher frequency interpulse and main pulse.} 
\label{fig_mean_profiles}
\end{center}
\end{figure}

We have focused on the Crab pulsar because its occasional very strong
``giant'' pulses are ideally suited to our data acquisition
systems. The mean profile of this star is dominated by a Main Pulse
(MP) and an Interpulse (IP), as shown in Figure
\ref{fig_mean_profiles}. Although the relative amplitudes and detailed
profiles of these features change with frequency, they can be
identified from low radio frequencies ($\ltw 300$ MHz) up to the
optical and hard X-ray bands.  The similarity of the mean profile
across this broad frequency range suggests that the radio emission and
high-energy emission arise from the same regions of the magnetosphere
in this star.

Several geometrical models have been proposed for the origin of the MP
and IP emission in pulsars. Traditional radio-pulsar models ascribe
MP/IP pairs to 
low-altitude emission (a few to a few tens of stellar radii) from
highly relativistic outflows above the star's two 
magnetic poles. Some models of high-energy pulsed emission  also 
locate the emission regions  at low altitudes (\emph{e.g.}, Daugherty
\& Harding 1996).  
If this is the case, the  magnetic axis of the Crab pulsar must be
nearly orthogonal to its rotation axis in order 
to see the highly beamed emission from both poles.  
Alternatively, some authors have suggested that the magnetic and rotation
axes are nearly aligned, and the MP and IP emission comes from a wide
emission cone (\emph{e.g.}, Manchester and Lyne 1977).  
Still other models relax constraints on the angle between the rotation
and magnetic axes, and locate both radio and high-energy emission sites  in
the outer magnetosphere,   
possibly  at the outer gap  described by  Cheng and Ruderman (1977).  
Yet another variant is the  caustic model of Dyks \emph{et al.}\
(2004), in which emission extends over a wide range of altitudes, from
the star's surface nearly to the light cylinder. (The Dyks \emph{et
  al.}\  model suggests   
IP emission comes from  higher altitudes than MP
emission, as does one of the models discussed by Hankins and Cordes
1981).

We do not know which, if any, of these models are correct,  but 
most of them suggest  physical conditions in the two emission
regions should be similar.  One  would expect the same radio emission
mechanism to be active in the IP and the MP.  
We were quite surprised,
therefore, to find  that the IP and MP have very different properties
at high radio frequencies (5-10 GHz), as we report in this paper.

\subsection{Observations and Post-processing}
In 2002 we captured strong, individual Crab pulses at the
Arecibo Observatory\footnote{The Arecibo Observatory is part of the
National Astronomy and Ionosphere Center, which is operated by
Cornell University under a cooperative agreement with the National
 Science Foundation.} at 2-ns time resolution, from $1.4\!-\!5$ GHz,
as reported in HKWE. For our new observations, reported here, we went
to higher frequencies, $6\!-\!8.5$ GHz and $8\!-\!10.5$ GHz, in order
to obtain 2.5-GHz 
bandwidth and consequent 0.4-ns time resolution.  For pulses that
exceeded a preset 
threshold the received voltages from both polarizations were digitized 
with 8-bit resolution and stored for off-line coherent dedispersion.
This allowed us to reach intrinsic time resolutions down to the limit
imposed by the inverse of the receiver bandwidth, 0.4 ns. In \S2 and
\S3 we show dedispersed individual pulses and their dynamic
spectra\footnote{The dynamic spectrum is computed from the dedispersed
voltage time series. It shows how the received pulse intensity is
distributed in time and radio frequency.} recorded at 8-10.5 GHz for all of our
figures; the results at 6-8.5 GHz are  similar in all
characteristics. 

After capturing a pulse our data acquisition system requires more than
a pulse period to store the data and be reset to capture another
pulse. Therefore we record only pulses which exceed a preset
threshold, which we set high enough to trigger the data acquisition
system only for the brightest individual pulses. The trigger detector
bandwidth was typically 0.5 GHz, centered on the 2.5-GHz sampled
bandwidth.

The pulses we record coincide with the high-flux power law tail
of the number-{\it vs.}-flux histogram for single pulses, as seen by Argyle
and Gower (1972) and Lundgren  \emph{et al.}\ (1995) at lower
frequencies;  thus they might be 
loosely called ``giant'' pulses.  However, it is not yet clear whether these
high-flux pulses are physically similar to, or different from, the
more common ``weak'' pulses;  we are not aware of any compelling
evidence for either case.   In what follows we do not attempt to
distinguish between the two, but just discuss MPs or IPs.

At the high time resolution we achieve, details of the pulses are
sensitive to the exact value of the dispersion measure (DM) used for
the coherent dedispersion operation. We generally started with the DM
value given by the Jodrell Bank Crab Pulsar
Monthly Ephemeris\footnote{www.jb.man.ac.uk/$\sim$pulsar/crab.html} for our
observing epochs. However, we found evidence in the dynamic spectra that
individual pulses could be more or less dispersed than the tabulated
value, and that IPs are systematically more dispersed than MPs, as
discussed in \S\ref{section_The_Interpulse}.  We then attempted to
find the ``optimum'' DM for most of the pulses we show in this
paper. For a narrow pulse, such as most components of a MP, we used
the DM value which maximized the peak intensity and the intensity
variance, minimized the width, and aligned the arrival times of
emission throughout our bandwidth. As shown in
\S\ref{section_The_Interpulse} and Figures \ref{typical_IP_1},
\ref{typical_IP_2} and \ref{typical_IP_3} 
the typical IPs are broader than the MPs; alignment of the dynamic
spectra provided more reliable DM estimates for the IPs. We found it
necessary to refine our optimum DM to a resolution of $10^{-5}$
pc-cm$^{-3}$.

In the remainder of this paper we present our observations of the main
pulse in \S\ref{section_The_Main_Pulse}, then the interpulse and its
narrow emission bands in \S\ref{section_The_Interpulse}. In
\S\ref{Emission_Band_models} we discuss some
possible causes of the interpulse emission bands and model
limitations, and summarize our results in
\S\ref{Conclusions}. 

\section{THE MAIN PULSE}

\label{section_The_Main_Pulse}
In our first high time resolution observations of the Crab pulsar
(reported in HKWE),  we concentrated on the MP, because it is
brighter at low frequencies, and strong pulses are more common at the
phase of the MP (Cordes \emph{et al.}\ 2004).  Our new observations at
higher frequencies, 6-8.5 GHz and 8-10.5 GHz at Arecibo, confirm and
extend our original results on the MP.   

\subsection{Microbursts and Nanoshots}

\begin{figure}[htb] % This is figure 2 (21 Oct 2006)
\begin{center}
\rotatebox{-90}{
\includegraphics[width=0.74\textwidth]{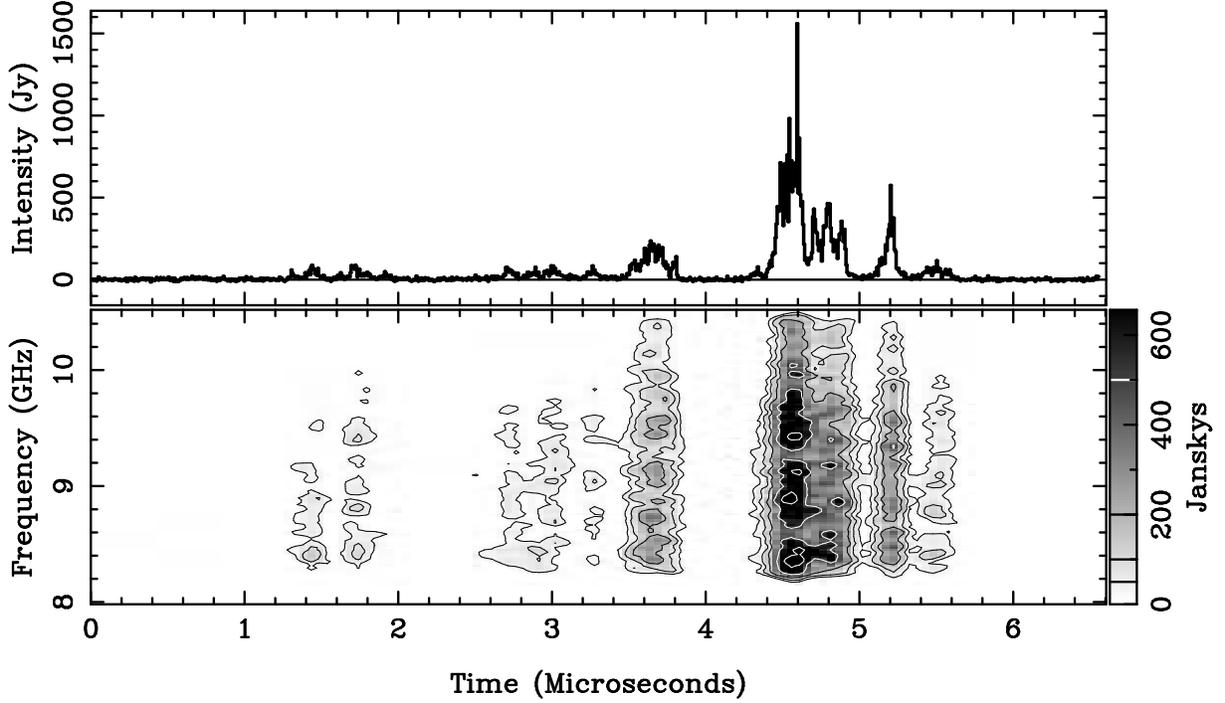}}
\caption{An example of a ``normal''  main pulse, processed with
  optimal dispersion measure, plotted with total intensity
  time resolution 6.4 ns, and dynamic spectrum resolution 51.2 ns,
  19.5 MHz.  The pulse seen in total intensity (upper panel) consists
  of several short-lived microbursts, which themselves contain
  shorter-duration nanoshots.  The dynamic spectrum (lower
  panel) reveals that the microbursts are broad-band, spanning the
  2.2-GHz receiver bandwidth.  The lack of emission at the lower band
  edge is because the 2.5-GHz sampled bandwidth is slightly larger
  than the receiver passband; the high-$\nu$ fading is intrinsic to
  the star. The spectrum contour levels are 50, 100, 200, 500, and
  1000 Jy. 
}
\label{fig_normal_MP_1}
\end{center}
\vspace{-2ex}
\end{figure}

\begin{figure}[htb] % 
\begin{center}
\rotatebox{-90}{
%\includegraphics*[width=0.66\textwidth,viewport=0 0 501 765]{f2.ps}}
% This is X20050809121338dynspec.0027-0027.ps
\includegraphics[width=0.74\textwidth]{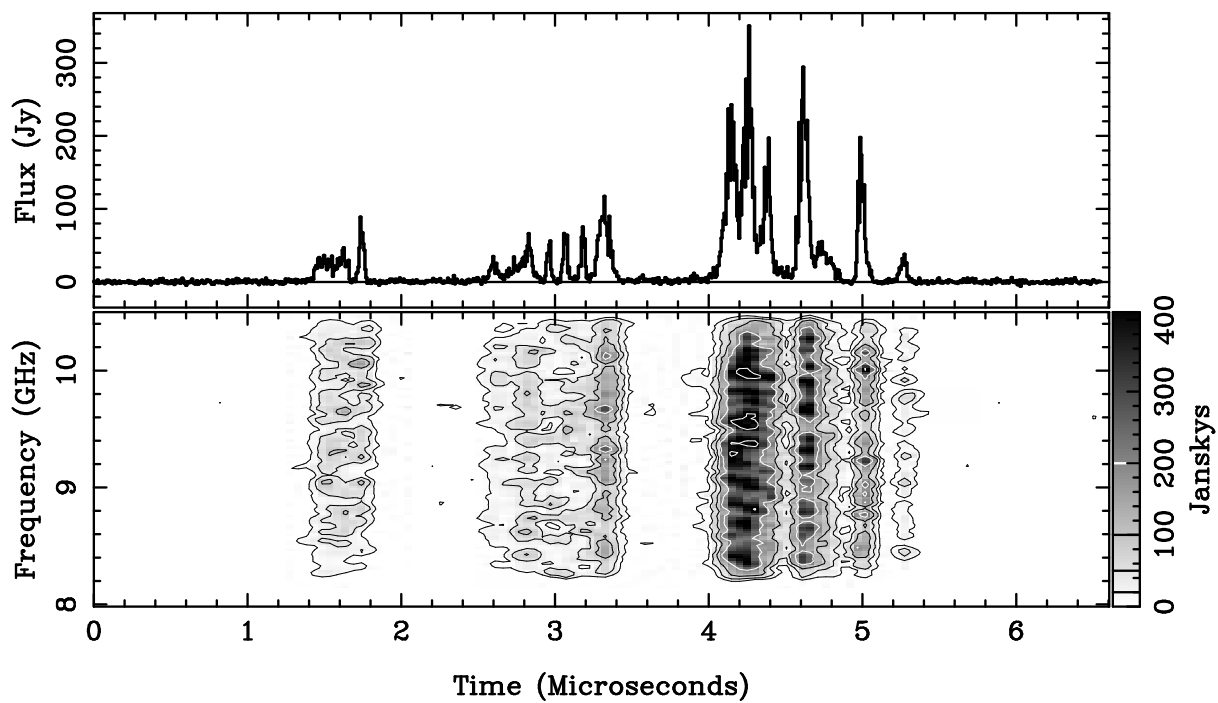}}
\caption{ Another example of a ``normal'' main pulse, also processed with 
optimal dispersion measure.  Similarly to the example in Figure
\ref{fig_normal_MP_1}, this pulse contains several short-lived
microbursts, each of which contains shorter-duration nanoshots.
This pulse differs from that in Figure \ref{fig_normal_MP_1} in
that its dynamic spectrum does not fade to high frequencies. Total
intensity time resolution 6.4 ns;  dynamic spectrum resolution
51.2 ns, 19.5 MHz.  Spectrum contour levels are 20, 50, 100, and
200 Jy.
}
\label{fig_normal_MP_2}
\end{center}
\vspace{-2ex}
\end{figure}

Most MPs consist of one to several ``microbursts'';  the
 brightest microburst in an MP can occur anywhere 
 within the pulse average envelope.    The microbursts can often be resolved into
overlapping, short-lived ``nanoshots''. 
Figures \ref{fig_normal_MP_1} and \ref{fig_normal_MP_2}
shows typical examples of  MPs; other examples are shown in
Sallmen \emph{et al.}\ (1999) and Kern (2004).

The  dynamic spectrum of the   microbursts is  broadband, filling our
entire observing bandwidth.  The emission is  sometimes, but not always,
slightly weaker toward the high frequency edge of the receiver band
(as illustrated in Figures \ref{fig_normal_MP_1} and \ref{fig_normal_MP_2}). 
This is   unlikely to be due to instrumental or interstellar
effects.  We normalized our system gain, as a function of frequency,
by the Crab Nebula background which dominates the off-pulse system temperature.
The nebular spectrum is quite  flat (proportional to $\nu^{-0.26}$  at
these frequencies;  Baars  \& Hartsuijker 1972). The  correlation
bandwidth due to interstellar scintillation (ISS) is predicted to be $\sim
1-2$ GHz at 8-10 GHz, and scales approximately as $\nu^4$ (Cordes
\emph{et al.}\ 2004, Kern 2004).  
We might expect interstellar effects occasionally to be seen in the
dynamic spectrum,  but such  effects should be equally likely to be
seen at the lower edge of the passband, as at the high edge.  We
therefore conclude that the high-frequency fading sometimes seen in the dynamic
spectrum of the MP is intrinsic to the MP emission mechanism.  This is 
consistent with the known steep radio spectrum of the Crab pulsar
($\nu^{-3.1}$;  Manchester, \emph{et al}. 2005).

\begin{figure}[htb]
\begin{center}
\rotatebox{-90}{
\includegraphics[width=0.74\textwidth]{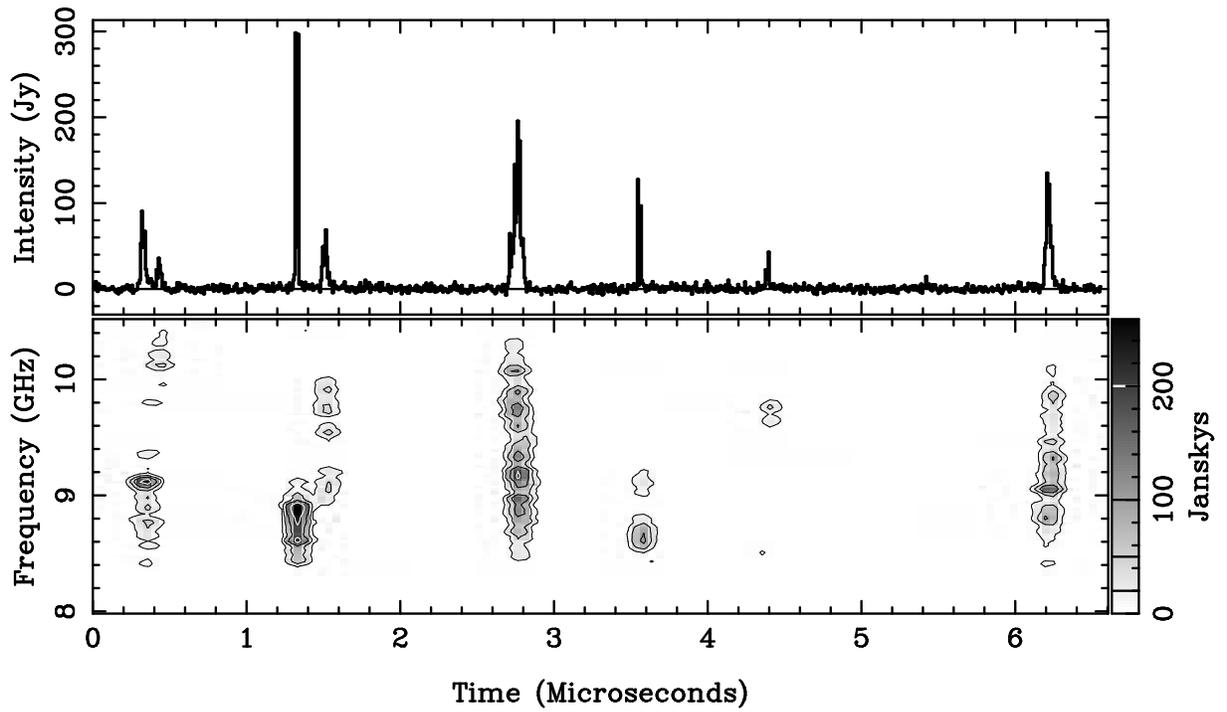}} 
\caption{An example of a sparse main pulse at 9.25 GHz. The pulse is plotted
  with the same total intensity and dynamic spectrum resolution as the
  pulse in Figures \ref{fig_normal_MP_1} and \ref{fig_normal_MP_2}.
 Occasionally nanobursts are
  sufficiently sparse that individual bursts with relatively narrow
  bandwidths can be identified, as seen here. The spectrum contour
  levels are 20, 50, 100, and 200 Jy.} 
\label{fig_sparse}
\end{center}
\vspace{-2ex}
\end{figure}

While most MPs resemble the examples in Figures \ref{fig_normal_MP_1}
and \ref{fig_normal_MP_2}, occasionally the nanoshots are sufficiently
sparse to be seen 
individually. Figure \ref{fig_sparse} shows one example.    When this
is the case, some of the nanoshots turn out to be relatively narrow-band.  We
believe these new data support our argument, in  HKWE, that each MP
microburst  is a collection of short-lived nanoshots.  When the 
time resolution is high enough, and the nanoshots are well separated
in time, the individual shots can be resolved.  We note that this picture is
consistent with previous modeling of pulsar emission as
amplitude-modulated noise, produced by the ensemble of a large number
of randomly occurring nanoshot pulses, modulated by a more slowly
varying amplitude function (Rickett 1975).

\subsection{Our Interpretation: Strong Plasma Turbulence }
\label{SPT theory}

We argued in HKWE that the nanoshots represent the fundamental
emission mechanism in  MPs.  In that paper we compared the
nanoshots to predictions of the three competing 
theoretical models of the radio emission mechanism. 
We found that the short durations and narrow bandwidths of
the nanoshots are consistent \emph{only}\/ with 
predictions of the SPT model.  They  are not consistent with
predictions of scaling arguments describing emission
from masers or from coherent charge bunches (both of which  predict
longer characteristic times).   In particular, Weatherall (1998)
modeled SPT and predicted narrow-band radiation, 
$\delta \nu / \nu  \sim 0.1-0.2$, centered on the co-moving plasma
frequency.  He also predicted a distinctive time signature, arising
from the coupling of the electromagnetic modes to the turbulence:
$\nu \delta t \sim O(10)$.   The relatively narrow-band spectra of the
nanoshots revealed by our new observations match Weatherall's models
well.  We thus confirm our suggestion, in HKWE, that coherent radio
emission in MPs is plasma emission produced by collapsing
solitons in strong plasma turbulence.

We note, however, that the SPT model makes no predictions on the spectrum
of a collection of nanoshots.  The spectral steepening we sometimes
see in the MP  dynamic spectrum could be due either to fewer nanoshots 
within a high-frequency microburst, or to less energy released in a
single high-frequency nanoshot, or both.

This model does make one important prediction:  plasma flow in the radio
emission region is likely to be  unsteady.  The  plasma
flow will be smooth only if the local charge density is 
exactly  the Goldreich-Julian (1969; ``GJ'') value, $\rho_{\rm GJ} \simeq
\Omega B / 2 
\pi  c$ (for a rotation rate $\Omega$), so that the rotation-induced
electric field is fully shielded. If  the charge density
differs from $\rho_{\rm GJ}$,  
the emitting plasma feels an unshielded electric field, and
feeds back on that field as its charge density fluctuates, leading to
unsteady plasma flow (and consequently unsteady 
radio emission). 

For the Crab pulsar, making the usual assumption that its spindown is
due to magnetic dipole radiation, we estimate a field $B(r) \simeq {4}
\times 10^{12} $ G, and a GJ density $n_{\rm GJ}(r) = \rho_{\rm GJ}(r)
/e \simeq {8} \times 10^{12}$ cm$^{-3}$, close to the star's surface.
At the light cylinder, $\sim 160 r_*$, the field drops to $\sim 1
\times 10^6$ G and the GJ density to $\sim 2 \times 10^6$ cm$^{-3}$.
Current pair cascade models predict that the (neutral) pair density
exceeds the primary beam (GJ) density by a factor $\lambda \sim
10^2\!-\! 10^3$ (Arendt \& Eilek 2002). 
Now, if SPT is the emission mechanism, we can determine the plasma density
directly, because SPT emission is centered on the comoving plasma
frequency ($\nu_p \propto \sqrt{\gamma_b n}$, for bulk Lorentz factor
$\gamma_b$).  Thus, emission at frequency $\nu$ comes from plasma
density $n \gamma_b \simeq 1.2 \times 10^{10} \nu_{\rm GHz}^2$
cm$^{-3}$.  Noting that current models predict $\gamma_b \sim
10^2-10^3$ ({\it e.g.}, Arendt \& Eilek), and using $\lambda$ to
convert from plasma density to charge density, our SPT argument
predicts that $1$-GHz emission comes from a region with number density
of excess charge $\sim 10^4 - 10^6$ cm$^{-3}$, and $9.5$-GHz emission
comes from $ \sim 10^6 - 10^8$ cm$^{-3}$. The higher density values
may be consistent with GJ conditions at moderate to high altitudes;
the lower values are very unlikely to satisfy GJ conditions anywhere
in the magnetosphere. We therefore suggest the highly unsteady, bursty
MP emission we see from 1 to 10 GHz reflects unsteady plasma flow due
to sub-GJ charge densities in the emission region (\emph{cf.}\ also
Kunzl \emph{et al.}\ 1998, who drew the same conclusion from a
somewhat different argument).

\subsection{Extreme Nanoshots}

\begin{figure}[htb]
\begin{center}
\rotatebox{-90}{
\includegraphics[width=0.65\textwidth]{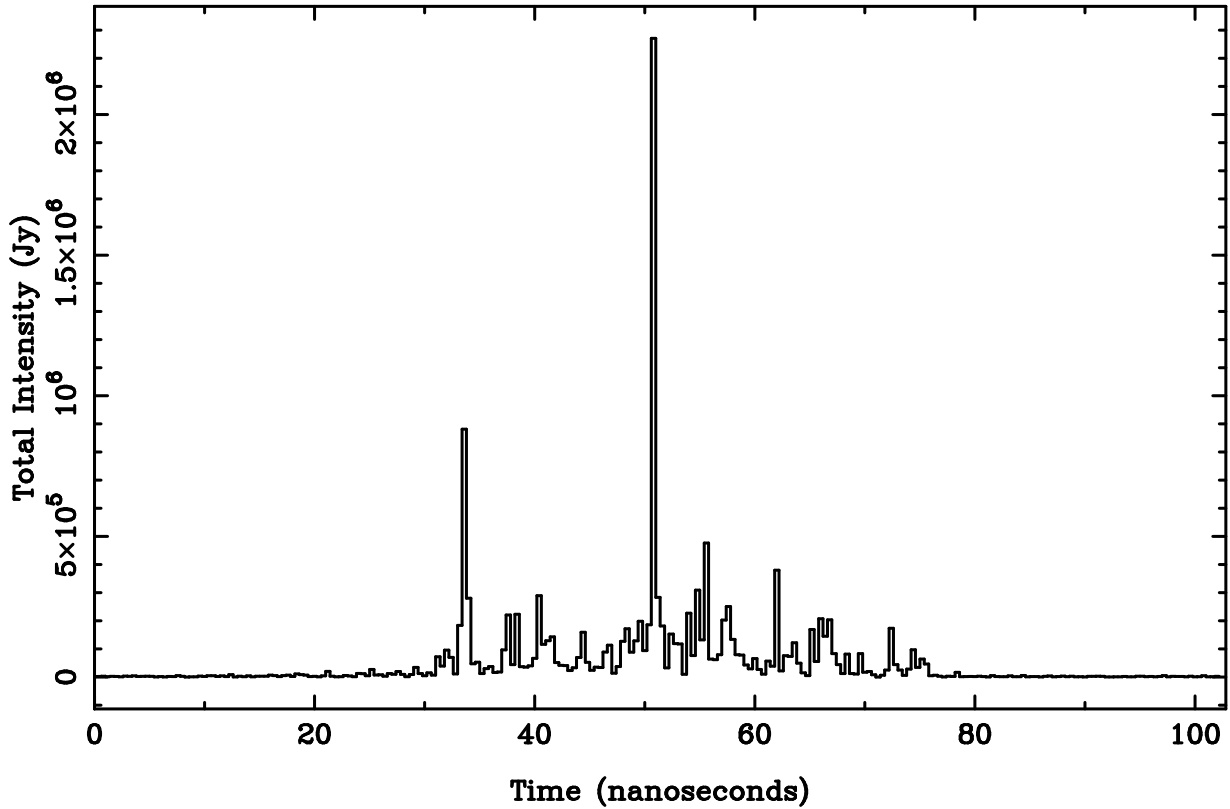}}
\caption{A single main pulse  recorded at 9.25-GHz center frequency over a
  2.2-GHz bandwidth and optimally dedispersed.  The nanopulse shown is
  unresolved with the 0.4-ns time resolution afforded by our
  system. Despite the high peak intensity of this pulse, it is
  unlikely that it saturated the data acquisition system. The
  dispersion sweep time across the bandwidth is about 1.5 ms, so as
  sampled by our data acquisition system, the dispersed pulse energy
  is spread over $\approx 7.5\times 10^6$ samples. }
\label{fig_2MJy_pulse}
\end{center}
\end{figure}

The nanoshots can occasionally be extremely intense. In Figure
\ref{fig_2MJy_pulse} we show a single MP which exceeds 2 MJy, and
has an unresolved duration of less than 0.4 ns. 
If we ignore relativistic effects (following, \emph{e.g.}, Cordes
\emph{et al.}\ 2004), we estimate a light-travel size $c \delta t  \simeq 12$ cm.
From this we find an  implied brightness temperature  $2 \times 10^{41}$K, 
which we believe is the highest ever reported for pulsar emission.
Alternatively, we might assume the emitting structure is moving
outwards with Lorentz factor $\gamma_b \sim 10^2 - 10^3$. 
If this is the case, our size estimate increases to $10^3 - 10^5$ cm,
and the brightness temperature decreases to $10^{35} - 10^{37}$K.

The extremity of this pulse can also be demonstrated in terms of
local quantities. If the pulse is emitted from a structure moving at
$\gamma_b$, its energy density is $u_{\rm  rad} \sim 4 \times
10^{23} / \gamma_b^4$ erg cm$^{-3}$. This high energy density can be
compared to the plasma energy density, $u_{\rm pl} = \gamma_b n m
c^2$, which we can estimate  either from our assumption of SPT
emission (which gives a lower density), or by assuming that GJ
conditions hold (which gives a higher density). In either case, we
find $u_{\rm rad}  \gg u_{\rm pl}$, unless $\gamma_b$ is extremely
large.  As noted in HKWE, this
emphasizes the need  for a collective emission process.   
For another comparison, we can  convert the energy density in the
radiation pulse to an equivalent electric field, $E \sim 3.2 \times
10^{12}/ \gamma_b^2 $G, giving a wave-strength parameter $e E / 2\pi m_e
c \nu \gg 1$. It follows that magnetospheric propagation of such
strong nanoshots  will be complex and nonlinear (\emph{e.g.}, Chian \&
Kennel 1983).

\section{THE INTERPULSE}
\label{section_The_Interpulse}
In order to test our hypothesis that strong plasma turbulence governs
  the emission physics in the Crab pulsar, we went to higher
  frequencies to get a larger bandwidth and shorter time  resolution.
  In addition to the MP, we observed single pulses from the IP, because
  at high frequencies strong pulses are far more common at the
  rotation phase of the IP. When we 
  observed IPs and MPs with a broad bandwidth, from 6-8 or
  8-10.5 GHz, we were astonished to find that IPs have very different
  properties from MPs.

\subsection{Characteristics}

We recorded about 220 individual MPs, and about 150
individual IPs,  between 
4.5 and 10.5 GHz, during 20  observing days from 2004 to 2006. 
Together with our earlier results (Moffett \& Hankins 1999), these
data reveal that the high-frequency IP  differs from the MP in
intensity, time signature, polarization, dispersion and spectrum.  In
this subsection we discuss the first four properties;  we defer
discussion of the spectrum to the next subsection.   We illustrate our
discussion with  Figures \ref{typical_IP_1}, \ref{typical_IP_2}
and \ref{typical_IP_3} which show three typical IPs. 
 
\subsubsection{Intensity}  Strong IPs are at least an order of
magnitude more frequent than strong MPs at 9 GHz,  but the MPs 
can be considerably stronger than the IPs 
when they occur.  This can be seen in the examples shown in this
paper, as well as in the the signal-to-noise-ratio histograms of
Figure 3 of Cordes \emph{et 
al.}\ (2004) and the scatter plots of their Figure 5 (where the
signal-to-noise-ratio of MPs and IPs are shown as a function of pulse
phase). 

\subsubsection{Polarization}  
High-frequency IPs are more strongly 
polarized than MPs. Moffett \& Hankins (1999) showed that the IP is strongly
linearly polarized,  50-100\% at 4.9 GHz, while the MP is only weakly polarized.
 We showed in HKWE that individual  nanoshots in the MP
can be strongly polarized, but the polarization changes dramatically
from one nanoshot to the next.  This leads to to weak MP polarization when
the nanoshot density is high or the pulse is smoothed to $\sim 1 \mu$s.

\subsubsection{Time signature}  IP emission is not broken up into the
short-lived microbursts that characterize the MP.  Instead, it is more
 continuous in time, spread out over a few microseconds.  When 
optimally dedispersed, IPs usually have a very rapid onset, followed
by a slower decay and often similar secondary bursts.  
To quantify the time duration of the IPs, we used the equivalent width of 
the intensity autocorrelation function to estimate the IP duration.
In Figure \ref{IP_widths} we show the distribution of equivalent widths for all
of the IPs we recorded at and above 6 GHz. This figure shows that IPs
typically last several microseconds at 6-8 GHz, and become shorter at higher
frequencies.

Because the temporal behavior of MPs is much more complex
(as we discussed in \S \ref{section_The_Main_Pulse}; {\it cf.} 
Figures \ref{fig_normal_MP_1} and \ref{fig_normal_MP_2}), the question
``what is the characteristic time signature of an individual MP'' is
difficult to answer.  We discuss MP time scales in a
forthcoming paper.

\subsubsection{Dispersion}  
\begin{figure}[htb]
\begin{center}
\rotatebox{-90}{
\includegraphics[width=0.74\textwidth]{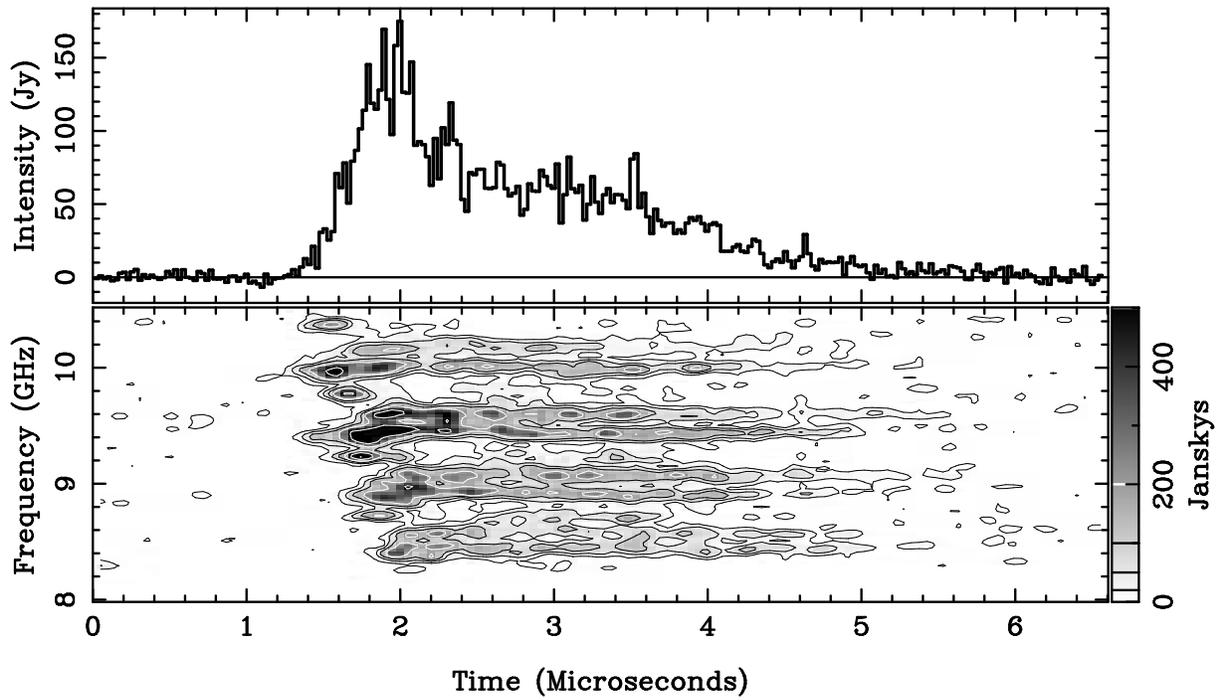}}
\caption{A typical interpulse, observed 12 minutes after
  the main pulse shown in Figure \ref{fig_normal_MP_1}, and processed
  identically. The later arrival of bands at lower frequency implies
  that this pulse is more dispersed than the main pulse in Figure
  \ref{fig_normal_MP_1}. The spectrum contour levels are 20, 50, 100,
  200, 500 Jy. Total intensity time resolution is 51.2 ns; dynamic spectrum
  resolution is 51.2 ns, 19.5 MHz.}
\label{typical_IP_1} 
\end{center}
\end{figure}

IPs are more dispersed than MPs measured at the same time.  
As an example, the  IP in Figure
\ref{typical_IP_1} was observed 12 minutes after the MP shown
in Figure \ref{fig_normal_MP_1}, and processed identically.  The dynamic
spectrum of the IP shows that lower frequencies arrive later than high
frequencies;  we take this as evidence for extra dispersion in the IP.
Because we consistently found IPs more dispersed than
MPs observed on the same day, we conclude this extra dispersion must
occur {\emph{in the pulsar's magnetosphere}}.

It is hard to compare the data to predictions of magnetospheric
dispersion, because we do not know the correct dispersion relation
for the magnetospheric plasma.  As a simple example, consider
dispersion from a cold, unmagnetized plasma.  The $0.65$-$\mu$s
delay between 8.4 and 10.4 GHz, for the pulse shown in Figure
\ref{typical_IP_1}, would correspond to an excess dispersion
measure, $\sim 0.032$ pc-cm$^{-3}$ ($\sim 10^{-3}$ of the total DM
measured by Jodrell Bank). If the magnetosphere were filled with
cold plasma at the GJ density, it would have a column density $\sim
1.3$ pc-cm$^{-3}$, far more than enough to account for the excess DM
of the IP.  But this estimate is naive.  Most of the magnetosphere
is strongly magnetized, and at low altitudes charges are constrained
to move only along field lines.  A more realistic dispersion law is
needed, but without knowing conditions through which the pulse
propagates it is not clear which law to choose.

 We discuss these complex issues more fully in a separate paper (Crossley
  \emph{et al.}\ 2007).

\subsection{Emission Bands}
\label{Emission_Bands}
The most striking difference between the interpulse and the main pulse
is found in the  dynamic spectrum.   An IP contains  microsecond-long
trains of  {\em emission bands}, as illustrated in Figures
\ref{typical_IP_1}, \ref{typical_IP_2} and \ref{typical_IP_3}.   
\emph{Every}\/ IP we have recorded
displays these emission bands.  However, MPs recorded during
the same observing sessions and processed identically do not show the bands.
The bands are, therefore, not due to instrumental or interstellar
effects, but are intrinsic to the star.

\subsubsection{Properties of the Emission Bands}

The emission bands are grouped into regular ``sets''; 2 or 3 band sets,
regularly spaced,  can usually be identified in a given IP. Individual
band sets last no more than a few microseconds.   All bands in a
particular set appear almost simultaneously, certainly to within
$\Delta t_{\rm start} < 100$ ns for optimally dedispersed pulses. This
requires that all of the bands must originate from a region no more
than $d=c\Delta t_{\rm start} < 30$ m in size.  
  
The IPs show a sharp onset, which is often associated with a very
short-lived ($\leq 100$ ns) band set. Additional  band sets often turn
on part way through the pulse, producing secondary bursts of
total intensity which also show the characteristic
fast-rise-slow-decay time signature.   Band sets that begin
later tend to last longer, up to the few-microsecond duration of the
total pulse.

\begin{figure}[htb]
\begin{center}
\rotatebox{-90}{
\includegraphics[width=0.74\textwidth]{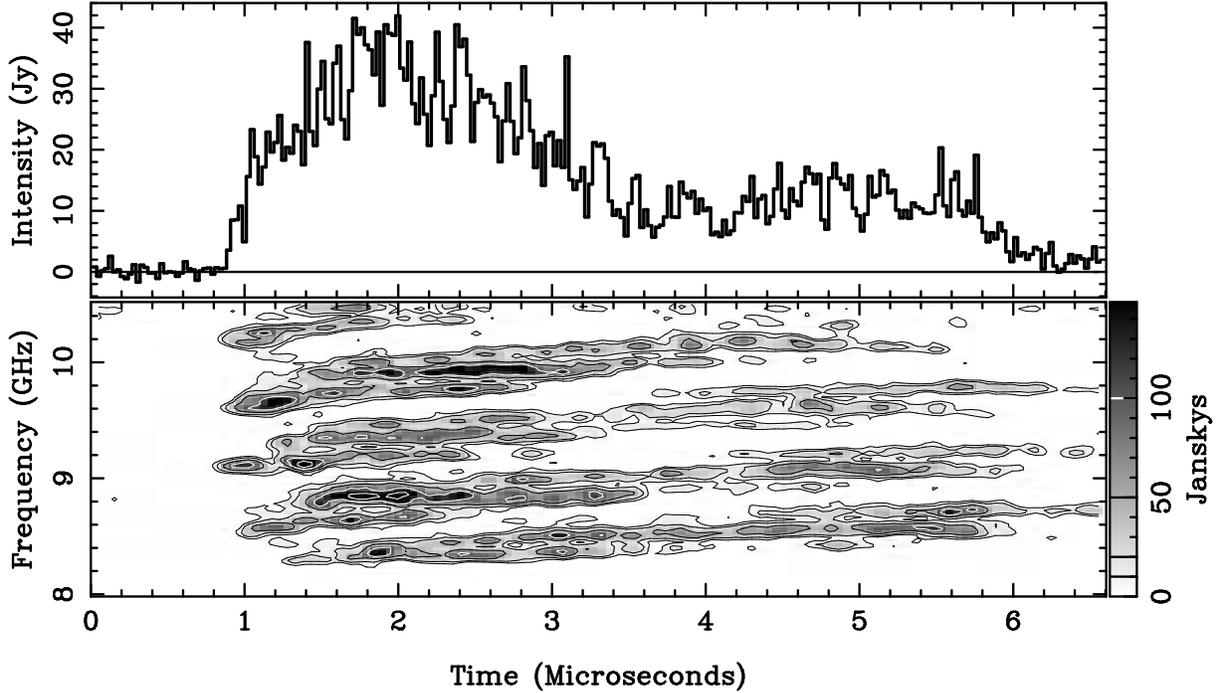}
}
\caption{Another interpulse, processed with the ``optimum'' dispersion
measure. All three of
the pulses shown in Figures \ref{typical_IP_1},
\ref{typical_IP_2} and \ref{typical_IP_3} show the emission bands we 
discovered in the dynamic spectrum of the interpulse. As these examples
show, several band sets can be identified within an interpulse. The interpulse
usually starts with a short-lived band set, and continues with
longer-lived sets, which either start at slightly higher frequencies
or drift upwards in frequency. The onset of later band sets often
coincides with a second burst in total intensity. As in Figures
\ref{fig_normal_MP_1} and \ref{fig_normal_MP_2}, the apparent lack of
low-$\nu$ mission is
because the sampled bandwidth is slightly larger than the receiver
bandwidth. Unlike the main pulse in Figure \ref{fig_normal_MP_1}, however, the
band intensity does not fade toward high frequencies. Here and in
Figure \ref{typical_IP_3}, the spectrum contour levels are 10, 20, 50, 100,
200 Jy. Total intensity time resolution is 51.2 ns; dynamic
spectrum resolution is 51.2 ns, 19.5 MHz. 
}
\label{typical_IP_2}
\end{center}
\end{figure}

\begin{figure}[htb]
\begin{center}
\rotatebox{-90}{
\includegraphics[width=0.74\textwidth]{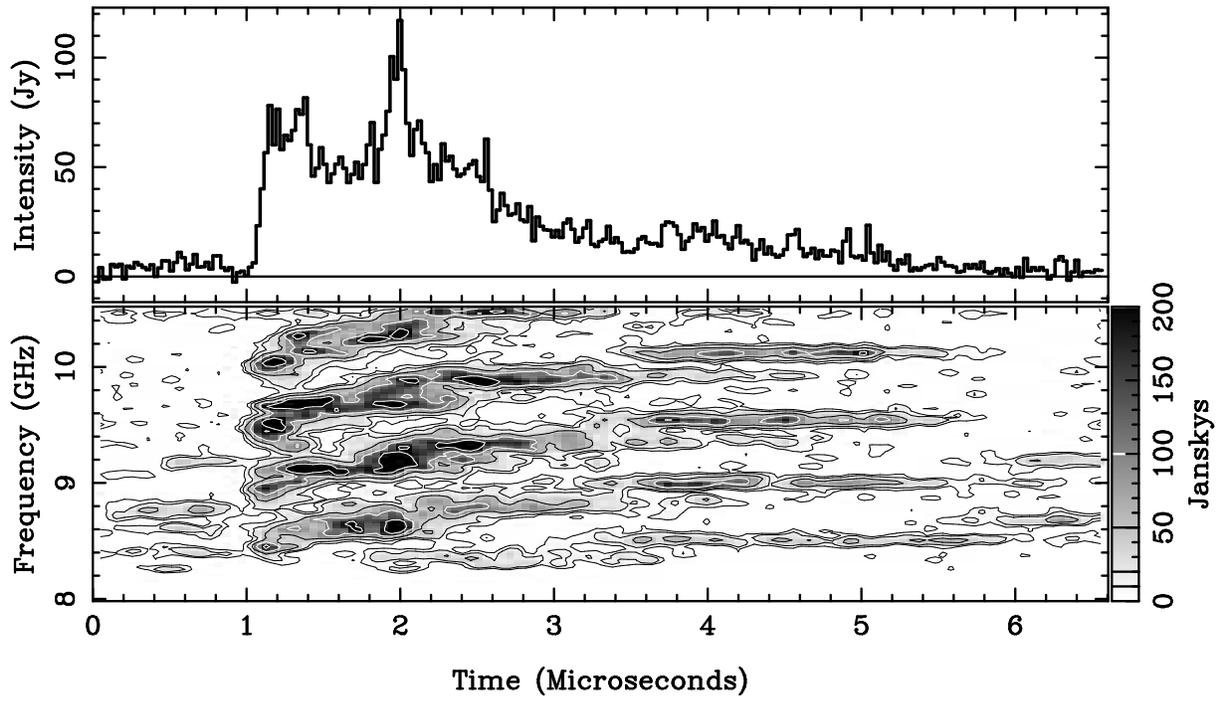}
}

\caption{Another example of an interpulse, also processed with ``optimum'' DM
  and the same time and spectral resolution as in Figures
  \ref{typical_IP_1} and \ref{typical_IP_2}. } 
\label{typical_IP_3} 
\end{center}
\end{figure}

\begin{figure}[htb]
\begin{center}
\rotatebox{-90}{
\includegraphics[width=0.725\columnwidth]{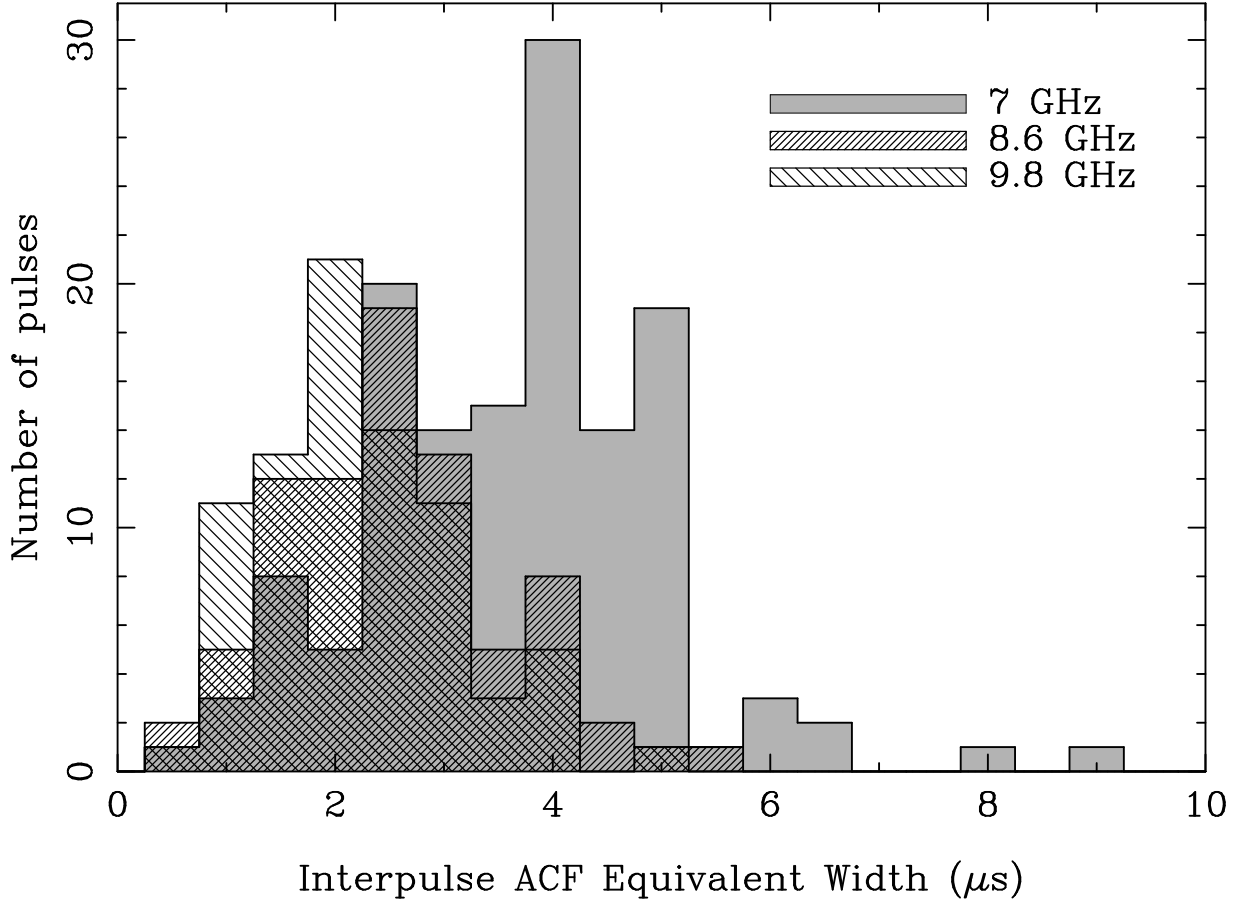}
}
\caption{The distribution of time durations of our interpulses, as
measured by the equivalent width of the autocorrelation
function. Nearly all of the interpulses captured at frequencies between 6 and 8
GHz were recorded with a 1-GHz bandwidth; for these pulses we used
the full bandwidth in computing the autocorrelation width. These
pulses are labeled as 7 GHz in the figure.  All of the interpulses between 8
and 10.5 GHz were recorded with a 2.5-GHz bandwidth.  For these we
divided the full band into high and low half-bands, each 1.25-GHz
wide, centered at 8.625 and 9.875 GHz,  and computed the autocorrelation width
separately for each half-band.   From the
distribution of time durations it is clearly seen that the interpulse
autocorrelation equivalent width is frequency dependent. The mean
widths are  4.2, 3.0, and 2.7 $\mu$s for 7, 8.6 and 9.9 GHz, respectively.
}
\label{IP_widths} 
\end{center}
\end{figure}

\begin{figure}[htb]
\begin{center}
\rotatebox{-90}{
\includegraphics[width=0.725\columnwidth]{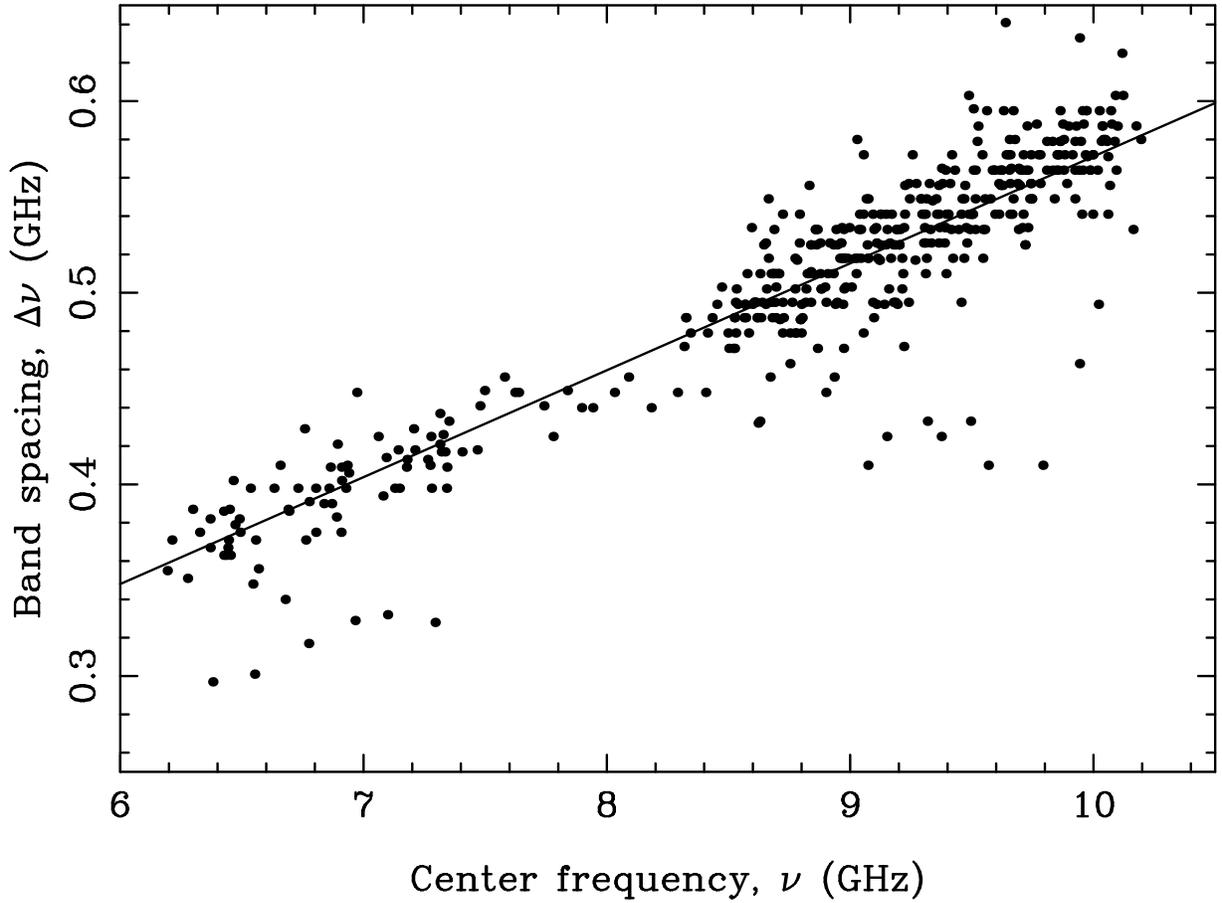}}
\caption{The emission band spacing, measured for 460 band sets in 105
  pulses recorded on 20 observing days, is shown as a function of
  center frequency. The line is fitted to all of the points, and has
  the form $\Delta \nu = 0.058(\pm 0.001)\nu - 0.007 (\pm 0.011)$. }  
\label{fig_band_freqs}
\end{center}
\end{figure}

At first glance the bands appear to be uniformly spaced.  However,
closer  inspection of our data shows that the bands are
\emph{proportionally   spaced}.  Figure \ref{fig_band_freqs} shows
that the  spacing between two adjacent bands depends on the mean
frequency,  as $\Delta \nu/ \nu \simeq 0.06$.  Thus, two bands near 6
GHz are spaced by $\sim 360$ MHz;  two bands 
near 10 GHz are spaced by $\sim 600$ MHz. This proportional
spacing is robust;  a set of emission bands can drift in frequency
(usually upwards, as in Figures \ref{typical_IP_1}, \ref{typical_IP_2}
and \ref{typical_IP_3}),   but 
their frequency spacing stays constant.  We note that the
least-squares fitted line in Figure  \ref{fig_band_freqs} is
consistent with zero spacing at zero frequency;  however we do not
think that the bands continue to very low frequencies (as we discuss in
the next section).  

The frequency profile of a given band tends to be peaked about a
central frequency, \emph{i.e.} closer to Gaussian than to rectangular or
impulsive.  The frequency width of an emission band, estimated by eye
as a half power width, is typically 10-20\% of the spacing between
bands.  Within a single IP the center frequency of an emission band
often remains steady until the band disappears.  In some instances,
however, the center frequency drifts upwards during the band duration,
by no more than 20-30\% of the band spacing.  We have occasionally
seen bands that appear to drift slightly downward in frequency, but
this is rare.  Bands sets that begin later in the pulse tend to start
at frequencies slightly higher than the early bands, but againw the
frequency shift is less than the spacing between bands.  These
features are all illustrated in Figures \ref{typical_IP_1},
\ref{typical_IP_2} and \ref{typical_IP_3}.  We note that quantitative
analysis of the structure of a single band is limited by the frequency
resolution we can achieve for short-lived bands, by signal-to-noise
limits and the overlap of separate band sets.

\subsubsection{Frequency Extent of the Bands}

We suspect the bands extend over at least a 5-6 GHz bandwidth in
a single IP, but do not occur below $\sim 4$ GHz. While we have not
been able to observe more than 2.2 GHz simultaneously, we have seen no
evidence that a given band set cuts off within our observable
bandwidth.  The characteristics of the bands (proportional spacing,
duration, onset relative to total intensity microbursts) are unchanged
from 5 to 10 GHz.   

We have captured a few IPs between 4 and 5 GHz, but the bands are
unclear in all of them. We suspect the dynamic spectrum in this
frequency range has been corrupted by ISS, for which the correlation
bandwidth is predicted to be $\sim 100$ MHz at these frequencies,
somewhat less than the band spacing projected from Figure
\ref{fig_band_freqs}.  Technical limitations, involving
terrestrial radio interference and the high-speed memory capacity of
our data acquisition system, kept us from observing with 
enough bandwidth to investigate the existence of bands
below $\sim 4$ GHz.  We note, however, that mean profiles suggest the
nature of the IP changes between 1 and 4 GHz.  From Figure 1 (also
Cordes \emph{et al.} 2004) we see that the IP between 4 and 8 GHz
appears at an earlier phase than at $\leq 1.4$ GHz, and there is no IP
at all around 3 GHz. This leads us to believe that the low and high
radio frequency IPs are unrelated;  we  therefore expect no IP band
emission below 4 GHz.

Unlike the MPs, there is no indication in our data
that the band intensity is weaker at high frequencies.   From this we
infer that the IP spectrum is flatter than the MP spectrum.  The IP
spectrum may, in fact,  be atypical of radio emission from the general
pulsar population,  which is known to be steep spectrum.

We reiterate that the emission bands in the IP cannot be due to ISS.
Their clear regularity, the fact that they exist only in the IP and
not in MPs observed at the same time, and their $\Delta \nu \propto
\nu$ spacing, all disagree with known properties of ISS, and point to
an origin intrinsic to the IP.

\section{POSSIBLE CAUSES OF THE EMISSION BANDS}

\label{Emission_Band_models} 
The dynamic spectrum of the interpulses does not match any of  the
three types of emission models described in
\S\ref{section_Introduction}.  Because each of  the models predicts
narrow-band emission at the plasma frequency,  \emph{none of them can
  explain the dynamic spectrum of  the IP}.  A new approach is
required here, which may ``push the envelope'' of pulsar radio
emission models.  
  
 While we remain perplexed by the dramatic dynamic spectrum of the
 interpulse, we have explored possible models.  This exercise is made
 particularly difficult by the fact that the emission bands are not
 regularly spaced. Because of this, models that initially seemed
 attractive must be rejected. As an example, if the emission bands
 were uniformly spaced they could be the spectral representation of a
 regular emission pulse train.  Many authors have invoked  regularly
 spaced plasma structures (sparks or filaments), whose passage across
 the line of sight could create such a pulse train. Alternatively,
 strong plasma waves with a characteristic  frequency will also create
 a regular emission pulse train.  The dynamic spectrum of either of
 these models would contain emission bands at constant spacing; the
 \emph{proportional}\/ spacing we observe disproves both of these
 hypotheses.  

We looked to solar physics for insight. We initially considered split
bands in  the dynamic spectra of Type II solar flares, which are
thought to be plasma emission from low and high density regions
associated with  a shock propagating through the solar corona.  This
does not seem to be helpful for the Crab pulsar  emission bands,
because the radio-loud plasma would have to contain 10 or 15 different
density stratifications, which seems  unlikely.  However, ``zebra
bands'' seen in Type IV solar flares may be germane.  These are
parallel, drifting, narrow emission bands seen in the dynamic spectra
of Type IV flares.  Band sets containing from a few up to $\sim 30$
bands  have been reported, with fractional spacing $\Delta \nu / \nu
\sim .01 - .03$ (\emph{e.g.}, Chernov \emph{et al.}\ 2005;  Sawant
\emph{et al.}\ 2005).  While zebra bands have not yet been
satisfactorily explained,  two classes of models have been proposed,
namely resonant plasma emission and geometrical effects.  It may be
that these models will provide clues to understanding the emission
bands in the Crab pulsar.  

\subsection{Resonant cyclotron emission}

One possibility is plasma emission at the cyclotron resonance, $\omega
- k_{\parallel} v_{\parallel} - s \Omega_0 / \gamma = 0$ (where
$\gamma$ is the particle Lorentz factor, $\Omega_0 = e B / m c$, and
$s$ is the integer harmonic number).  Kazbegi \emph{et al.}\ (1991)
proposed that this resonance operates at high altitudes in the pulsar
magnetosphere, possibly generating both X-mode waves (which can escape
the plasma directly) and O-mode waves; a related model is being
applied to the Crab emission bands by Lyutikov (2007, private
communication). Alternatively, double resonant cyclotron emission at
the plasma resonant frequency has been proposed for zebra bands in
solar flares (\emph{e.g.}, Winglee \& Dulk 1986).  In solar
conditions, this resonance generates O-mode waves, which must mode
convert in order to escape the plasma.  The emission frequency in
these models is determined by local conditions where the resonance is
satisfied; the band separation is $\Delta \nu \simeq \Omega_0 / 2 \pi
\gamma$.  It is not clear how the specific, proportional band spacing
can be explained in such models; perhaps a local gradient in the
magnetic field can be invoked.

Such models face several additional challenges before they can be
considered successful.  The emission must occur at high altitudes, in
order to bring the resonant (cyclotron) frequency down to the radio
band.  Close to the light cylinder, where $B \sim 10^6$ G, particle
energies $\gamma \sim 500 - 1000$ could radiate at 5 - 10 GHz. In
addition, such models must be developed by means of specific
calculations which address the fundamental plasma modes and their
stability, under conditions likely to exist at high altitudes in the
pulsar's magnetosphere.

\subsection{Geometrical models}
   Alternatively, the striking regularity of the bands calls to mind a
  special geometry.  If some mechanism splits the emission beam
  coherently,  so that it interferes with itself, the bands could be
  interference fringes.  For instance, a downwards beam which reflects
  off a high density region could return and interfere with its
  upwards counterpart on the way back up (\emph{e.g.}, Ledenev
  \emph{et al.}\ 2001 for solar zebra bands).  Simple geometry
  suggests that fringes occur if the two paths  differ in length by
  only $c / \Delta \nu \ltw 1$ m. Another geometrical possibility is
  that cavities form in the plasma and trap some of the emitted
  radiation, imposing a discrete frequency structure on the escaping
  radiation   (\emph{e.g.},   LaBelle \emph{et al.}\ 2003).  The
  scales required   here are also small;  the  cavity scale must be
  some multiple of   the wavelength, \emph{i.e.}, several centimeters.

Geometrical models need an underlying broad-band radiation source,
with at least 5-GHz bandwidth, in order to produce  the emission bands
we observe.  Because standard pulsar radio emission mechanisms lead to
relatively narrow-band radiation, at the local plasma frequency, they
seem unlikely to work here.  A double layer might be the radiation
source; charges accelerated within the layer should radiate broadband,
up to $\nu \sim L / 2 \pi c$, if $L$ is the thickness of the double
layer (\emph{e.g.}, Kuijpers 1990;  Volwerk  1993). This is
also a small-scale effect;  emission at 10 GHz requires $L \sim 1$ cm.
This may be consistent with the thickness of a relativistic, lepton
double layer, which we estimate as several  times $c / 2 \pi \nu_p$
(following, \emph{e.g.},  Carlqvist 1982).  

Geometrical models also face several challenges before they can be
considered successful. The basic geometry is a challenge: what
long-lived plasma structures can lead to the necessary interference or
wave trapping?  Once again, it is not clear how the proportional band
spacing can be explained;  perhaps a variable index of refraction in
the interference or trapping region can be invoked.  

\section{Summary and conclusions}
\label{Conclusions}
We have observed individual pulses from the Crab pulsar with 2.2-GHz
bandwidth and 0.4-ns time resolution. We observed both the main pulse
(MP) and the interpulse (IP), at high radio frequencies ($5\!-\!10$
GHz).  We were very surprised when our observations revealed that the
MP and the IP  are strikingly different.   
At these frequencies, MPs consist of many short-lived, relatively
narrow-band,  nanoshots.  Both the time and frequency signatures of
the nanoshots are consistent with  predictions of one current model of
pulsar radio emission, namely, coherent emission from strong plasma
turbulence.   

IPs, however,  differ from MPs in their time, polarization,
dispersion   and spectral signatures. It seems that the MP and the IP
differ in their emission mechanisms and their propagation within the
magnetosphere. The dynamic spectrum of the IP contains regularly
spaced emission bands, which do not  match the predictions of
any current model. Our  result is especially surprising because
magnetospheric models generally ascribe the MP and the IP to
physically similar regions, which simply  happen to be on opposite
sides of the star.  One would therefore expect the MP and the IP to
have similar characteristics, which is exactly \emph{not} what we find.   

The work of Moffett \& Hankins (1996) may provide an important
clue. They discovered new components in the mean profile at 5 and 8
GHz, which appear at offset rotation phases, and only at high radio
frequencies.  They also found that the IP at high radio frequencies
appears at an earlier rotation phase than its counterpart at both
lower (radio below 2 GHz) and higher (optical to X-ray) frequencies.
It is just this high-radio-frequency IP which we discuss in this
paper.  We speculate that this IP, and the other high-radio-frequency
components Moffett \& Hankins found, originate in an unexpected part
of the star's magnetosphere, where different physical conditions
produce quite different radiation signatures.

\begin{acknowledgements}

We appreciate helpful conversations with Joe Borovsky, Alice Harding,
Axel Jessner,  Jan Kuijpers, Maxim Lyutikov, and the members of the
Socorro pulsar group.  We thank our referee for constructive comments
which improved the paper.   We particularly thank our students, Jared
Crossley, Eric Plum and James Sheckard for their help with
observations. This work was partially supported by the National
Science  Foundation, through grants AST0139641 and AST0607492. 
\end{acknowledgements}

%%%%%%%%%%%%%%%%%%%% Figures and Captions %%%%%%%%%%%%

%%%%%%%%%%%%%%%%%%%%%%%%%%%%%%%%%%%%

%%%%%%%%%%%%%%%%%%%%%%%%%%%%%%%%%%%%
%%%%%%%%%%%%%%%%%%%%%%%%%%%%%%%%%%%%%%

%%%%%%%%%%%%%%%%%%%%%%%%%%%%%%%%%%%%%%

%%%%%%%%%%%%%%%%%%%%%%%%%%%%%%%%%%%%%%

%%%%%%%%%%%%%%%%%%%%%%%%%%%%%%%%%%%%%%

%%%%%%%%%%%%%%%%%%%%%%%%%%%%%%%%%%%%%%

%%%%%%%%%%%%%%%%%%%%%%%%%%%%%%%%%%%%%%

%%%%%%%%%%%%%%%%%%%%%%%%%%%%%%%%%%%%%%

\end{document}